\documentclass[journal=jacsat,manuscript=article]{achemso}

\usepackage[version=3]{mhchem} 
\usepackage[dvipsnames]{xcolor} 

\author{Caique C. Oliveira}
\affiliation[Federal University of ABC]
{Center for Natural and Human Sciences (CCNH), Federal University of ABC (UFABC), 09210-170, Santo André-SP, Brazil}

\author{Pedro A. S. Autreto}
\affiliation[Federal University of ABC]
{Center for Natural and Human Sciences (CCNH), Federal University of ABC (UFABC), 09210-170, Santo André-SP, Brazil}
\email{pedro.autreto@ufabc.edu.br}

\title[An \textsf{achemso} demo]
  {Strain Modulated Catalytic Activity of $Pt_{2}XSe_{3}$ (X = Hg, Zn) for Hydrogen Evolution Reaction}

\abbreviations{IR,NMR,UV}

\begin{document}




\begin{abstract}
The catalytic properties of $Pt_2XSe_3$ (X = Hg, Zn) in hydrogen-electrode- (HER-) based catalysts have been investigated based on state-of-the-art ab initio simulations. Our results show that the late transition metal sites (Hg and Zn) exhibit the best activity for HER in an acidic environment. Furthermore, lattice stretching and compression can effectively modulate the H binding energy, achieving almost thermoneutral adsorption at $3\%$ compressive strain. The changes are attributed to the modulation in the d-band centers of late transition metal sites, as well as the depletion of charge population on bonding states, contributing to the destabilization of the H-metal bonds. Our contribution explores strain engineering as an effective strategy to tailor the activity of 2D mineral-based catalyst materials for HER, advancing our understanding of how mechanical manipulation can effectively modulate the catalytic properties of these materials.
\end{abstract}

\section{Introduction}


The increasing consumption of energy accentuates the imperative for clean, renewable, and efficient energy sources as viable alternatives to the diminishing reserves of fossil fuels that predominantly govern the global energy matrix \cite{owid-electricity-mix, british2022british}. In this context, the advancement of strategic technologies, including batteries, fuel cells, and electrolyzers, for the purpose of energy production and conversion, is critically important to facilitate the decarbonization of essential energy sectors \cite{RAMASUBRAMANIAN2024, ABDIN2024111354, AKYUZ2024ELECTROLYZERS}. Hydrogen emerges as one of the most promising solutions owing to its high energy-to-mass ratio and its versatility for renewable energy production and storage \cite{AFONIN201611199}. Green hydrogen, produced by water electrolysis employing clean and renewable energy sources (such as wind and solar) in the Hydrogen Evolution Reactions [REF], has attracted significant interest due to its inherently sustainable nature. High-performance electrolyzers typically utilize noble metal-based catalysts, notably platinum, which restricts their commercial viability because of the limited availability of these materials. Thus, minimizing the noble metal content is crucial in the development of economically viable catalysts.

Two-dimensional materials have been extensively explored for their potential in catalysis applications \cite{SHANMUGHAN202328679, YANG2024, Zhang2DAng2025}. The high surface area and enhanced charge mobility facilitate electron transfer, thereby augmenting their catalytic properties. Transition Metal Dichalcogenides (TMDs) have similarly been investigated extensively within this context \cite{D3TA04475K}. For hydrogen evolution reactions (HER), it has been previously evidenced that the activity is more pronounced at edge sites as opposed to the basal plane \cite{He2020,ZHAOTMDsHER2025}. Conversely, the basal planes of polymorphic 1T TMDs exhibit greater catalytic activity compared to the more typical 2H phases \cite{TSAI2015133, LI2023107119}. Furthermore, doping, defect creation, as well as phase and strain engineering, are established strategies that can effectively modulate the electronic structure of these materials, promoting their catalytic properties \cite{ZHAOTMDsHER2025}. Notably, strain engineering is among the most prevalent methodologies applied to tailor the catalytic activity of 2D TMDs. Lattice expansion or contraction can be achieved through the application of curvature \cite{Jiang2021} or via the formation of heterostructures with materials exhibiting slightly disparate lattice parameters \cite{Yang2025}. Yang and colleagues have explored the effects of strain on the electronic, catalytic, and optical properties of $MoS_{2}/ZnO$ heterojunctions \cite{Yang2025}. Lee and his team demonstrated that tensile strain could induce thermoneutral hydrogen binding in $PtSe_{2}$ flakes \cite{JeonPtS2strain2025}. Furthermore, Lee and collaborators reported on the phase transformation in $MoTe_{2}$ from the 2H to the 1T phase, which results in a metallic character with enhanced catalytic activity \cite{LeeMoTe22023}. 

\begin{figure}[t!]
\centering
\includegraphics[scale=0.7]{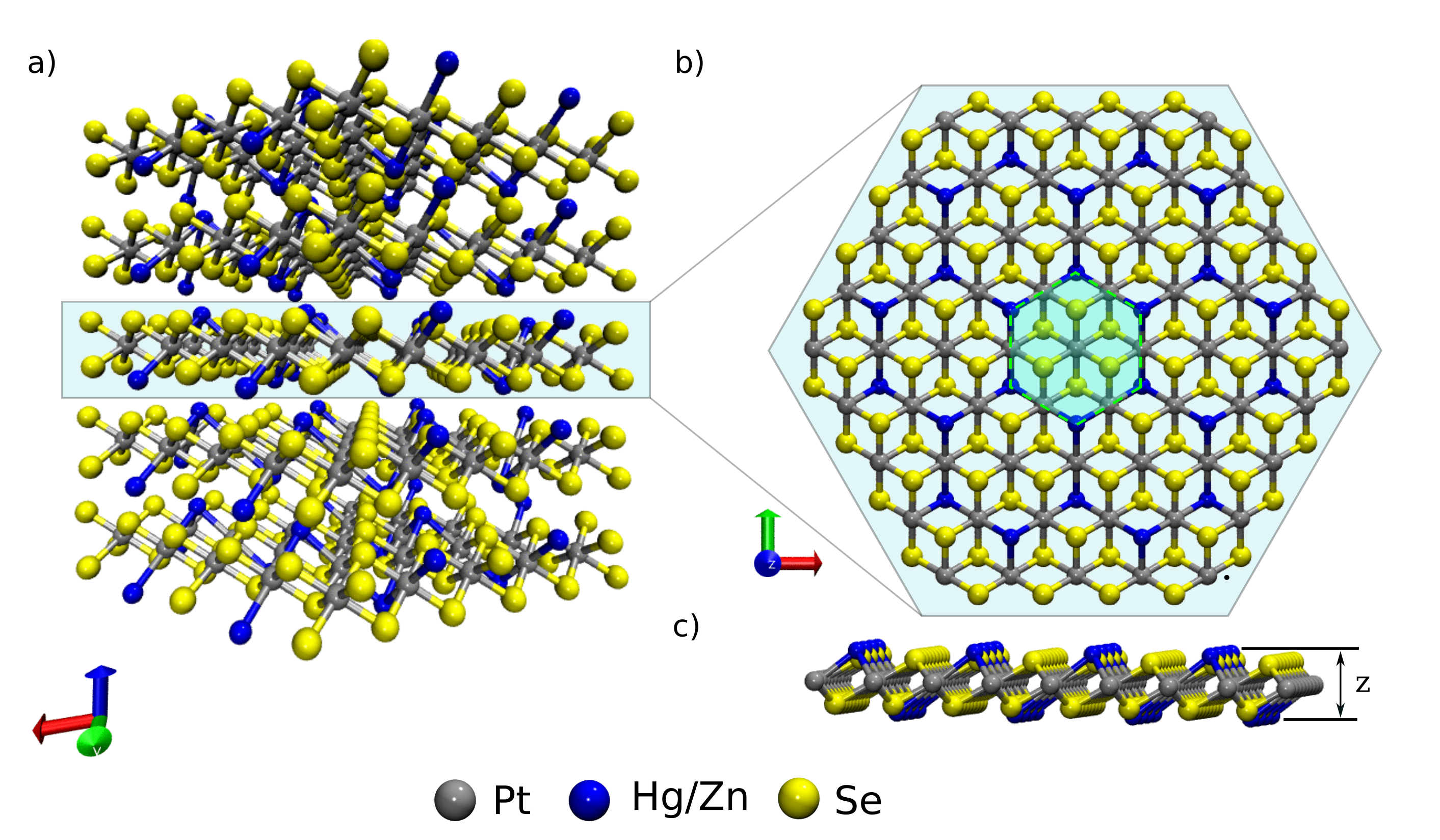}
 \caption{(a) Jacutingaite structural model composed by $Pt_{2}NSe_{3}$ (N = Hg, Zn) stacked by van der Waals forces as shown in. In each monolayer one fourth of the chalcogens are substituted by an N atom producing the hexagonal patterns shown in (b). The N atoms exhibit different buckling heigh (z) compared to Se as shown in (c).}
\label{FIG:1}
\end{figure}

Jacutingaite (chemical formula $Pt_{2}HgSe_{3}$, Figure \ref{FIG:1}.a) \ref{FIG:1}.a, is a naturally occurring compound initially isolated from the Caue mine in Itabira, Minas Gerais, Brazil \cite{CabralTerraNova2008, Vymazalova2012}. In addition to its natural occurrence, this compound has also been previously synthesized \cite{Longinhos2020, Cucchi2020, Kandrai2020}, which has garnered significant interest. Lima et al. have methodically examined the stability of the Jacutingaite family, a group of materials characterized by the general formula $M_{2}XN_{3}$, where M represents Pt, Pd, and Ni, X denotes Zn, Hg, and Cd, and N corresponds to S, Se, and Te, all of which exhibit high stability \cite{deLimaPRB2020}. Despite the growing interest in the topological characteristics of Jacutingaite structures, a comprehensive assessment of their potential for energy storage and conversion applications remains absent from the current literature. Consequently, in this study, we examine the effects of strain on the catalytic properties of $Pt_{2}XSe_{3}$ (X = Hg, Zn) in the context of hydrogen evolution reactions.

\section{Methods}

Spin-polarized first-principles calculations were carried out using the \textit{Vienna Ab-initio Simulation Package} (VASP) \cite{KresseVASP1996}. Electron-ion interactions were approximated using the Projected Augmented Wave (PAW) method \cite{KressePAW1999}. Exchange and correlation interactions were addressed through the Perdew, Burke, and Erzerhoff parameterization of the generalized gradient approximation \cite{pbe}, incorporating nonlocal van der Waals interactions corrections as implemented in the DFT-D3 functional \cite{Grimme2006}. Kohn-Sham orbitals were expanded using a plane-wave basis set, with a kinetic energy cutoff of 400 eV. The sampling of the Brillouin Zone (BZ) was conducted on a uniform k-point grid in accordance with the Monkhorst and Pack scheme \cite{Monkhorst}. Structural optimization was carried out with a 2x2x1 grid. For electronic structure calculations, which require refined sampling of the Fermi surface in the BZ, an 8x8x1 grid was employed. Throughout the simulations, a self-consistency threshold of $10^{-7}$ was maintained, with forces minimized below $0.02$ eV/\AA. Subsequent data post-processing was performed using the VASPKIT suite \cite{VASPKIT}. The Crystal Orbital Hamilton Population (COHP) analysis \cite{Dronskowski1993} was carried out to investigate the bonding nature of the H intermediate sites using the LOBSTER package \cite{Maintz2016}.

The catalytic activity of the structures was accessed by computing the H adsorption free-energy ($\Delta G_{H^{*}}$) employing the Computational Hydrogen Electrode (CHE) model of Norskov and coworkers \cite{Nørskov}, calculated as follows:
\begin{align}
    \Delta G_{H^{*}} = E_{H^{*}} - \left( E_{*} + \frac{1}{2}E_{H_{2}} \right) + \Delta E_{ZPE} - T\Delta S
\end{align}
where $E_{H^{*}}$ represents the total energy of the structure with an adsorbed H,  $E_{*}$ the total energy of the bare structure, $E_{H_{2}}$ the total energy of a $H_{2}$ molecule (in gas phase, $p_{H_{2}} = 1$ bar) and T the absolute temperature (298,15 K employed in this work). $\Delta E_{ZPE}$ and $\Delta S$ are the zero-point energy and entropic variation with respect to $H_{2}$ gaseous phase. Following previous works, these last two terms are approximated by 0.24 eV \cite{Nørskov_2005}. In this scheme, good catalysts should exhibit thermoneutral H binding with $\Delta G_{H} \approx 0$.

\begin{table}
\caption{Optimized lattice parameters and buckling height for $Pt_{2}XSe_{3}$}\label{tab:lattice-parameters}
	\centering
	\begin{tabular}{cccc}
        Structure & $a$ (\AA) & $b$ (\AA) & z (\AA) \\ \hline
        $Pt_{2}HgSe_{3}$ & 7.51 & 7.51 & 3.50 \\
        $Pt_{2}ZnSe_{3}$ & 7.46 & 7.46 & 2.72 
        \end{tabular}
\end{table}

\section{Results and discussion}

\begin{figure}[t!]
\centering
\includegraphics[scale=0.50]{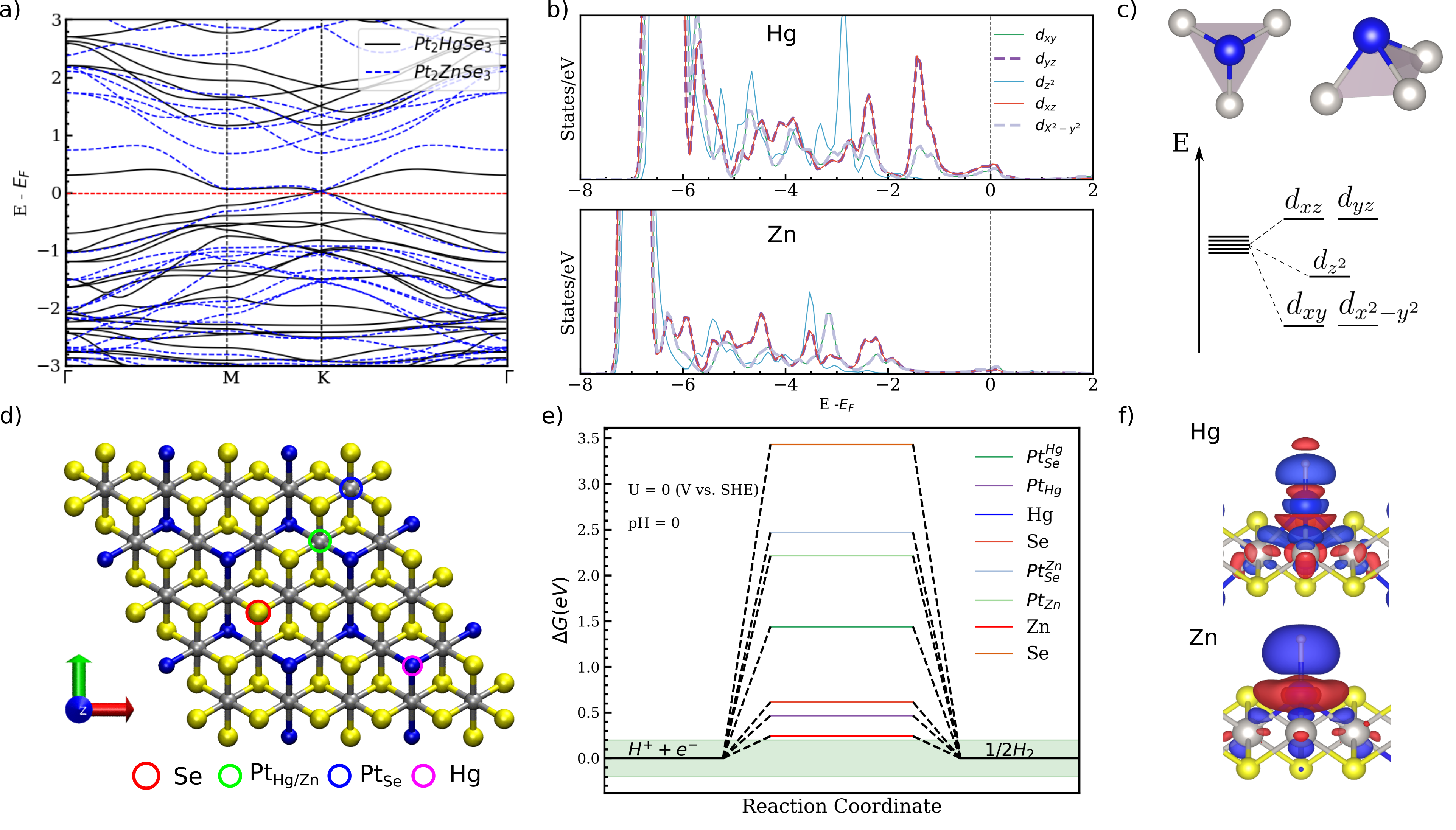}
 \caption{a) The band structure for $Pt_{2}HgSe_{3}$ (black) and $Pt_{2}ZnSe_{3}$ (dashed blue). b) Projected Density of States (pDOS) for Hg and Zn c) Illustration of the proposed d orbital splitting for Hg and Zn on trigonal coordination d) Top view of 2x2 supercell of $Pt_{2}XSe_{3}$ highlighting the H adsorption sites. e) Free-energy diagrams for H adsorption and f) Charge difference plot for H adsorbed on the Hg (top) and Zn (bottom) sites. The red (blue) isosurfaces show charge accumulation (depletion). Both isosurfaces were set to $1.25 \times 10^{-3}$ e/\AA$^{3}$}
\label{FIG:2}
\end{figure}

Jacutingaite is a van der Waals-layered compound, structurally similar to sudovikovite ($PtSe_{2}$) \cite{Vymazalova2012}. The structure is modified in such a way that one-fourth of the chalcogen atoms are substituted by a late transition metal: Zn, Hg, or Cd, forming a sublattice of hexagons as shown in the top view of a monolayer depicted in Figure \ref{FIG:1}.b. Therefore, the platinum (Pt) atoms can be either coordinated with six selenium (Se) atoms or coordinated with four selenium and two late metal atoms (Hg or Zn). The different coordination gives rise to a buckling height ($z$), since the bond lengths are expected to change in Pt-Se and Pt-Hg/Zn-Se octahedral. The optimized lattice parameters are a = b = 7.50 \AA with a buckling height (z) of 3.50 \AA for $Pt_{2}HgSe_{3}$ and a = b = 7.46 \AA with z = 2.72 \AA for $Pt_{2}ZnSe_{3}$, in good agreement with previous studies \cite{deLimaPRB2020, Santos-Castro_2025}. The electronic band structure is shown in Figure \ref{FIG:2}.a, from which it can be seen that the modification of the X species does not change the overall electronic behavior of the structure. In other words, the non-zero gap character is conserved, and the Dirac-like feature at the $K$ point is preserved. This result is in good agreement with previous calculations using DFT \cite{deLimaPRB2020}. Analyzing carefully, one can notice that the behavior of the valence and conduction bands is similar near the $K$ high-symmetry point, and the dispersion relation becomes very different towards the $\Gamma$ point: for Zn it seems that the band broadens in this direction. Furthermore, for Hg there is a gap in the conduction region right above the first conduction band whereas for Zn this gap is smaller. Generally speaking, if one disregards the last (first) valence (conduction) bands, when X = Zn, the bands shift to lower energy values. The projected density of states (pDOS) for the X atom is shown in Figure \ref{FIG:2}.b, indicating the degeneracy of the out-of-plane orbitals. The $d_{xy}$ and $d_{x^{2}-y^{2}}$ orbitals superpose each other, as well as the $d_{xz}$ and $d_{yz}$ orbitals. However, since no ligands are on the x and y plane, the $d_{xy}$ and $d_{x^{2}-y^{2}}$ orbitals should exhibit less electrostatic repulsion and therefore are expected to be in lowest energy state. On the other hand, the $d_{xz}$ and $d_{yz}$ present increased repulsion given the distorted trigonal character of the coordination environment. Thus, these two orbitals should be in at highest energy state relative to the others. The observed distorted trigonal coordination of the Hg/Zn atoms in the structure causes the $d_{z^{2}}$ orbital to be in lower energy state compared to $d_{xy}$ and $d_{x^{2}-y^{2}}$, however, since the ion is dislocated on the z axis, the intermediate repulsion should cause the $d_{z^{2}}$ orbital to be in an intermediate state, resulting in the d orbital splitting illustrated in Figure \ref{FIG:2}.c \cite{PhysRevMaterials.5.054001}.

The catalytic activity of both structures was investigated by calculating the H adsorption free-energy ($\Delta G_{H^*}$) for the non-equivalent sites of the $Pt_{2}XSe_{3}$ monolayers. In total, four sites were considered in each structure: the top Se sites and the top late transition metal (Hg or Zn) sites. Also, the two distinct coordination environments for Pt have been taken into account: $Pt_{Se}$ sites represent the case where the noble-metal is coordinated with 6 Se, whereas the $Pt_{Hg/Zn}$ corresponds to the octahedral containing two X and 4 Se atoms. A representative of each site is illustrated in Figure \ref{FIG:2}.d. The corresponding $\Delta G_{H^*}$ values are presented in Table \ref{tab:delta-G} and the resulting free-energy diagrams are shown in Figure \ref{FIG:2}.e, where it can be readily seen that only the N sites (Hg or Zn) shows promising activity with $+0,24$ and $+0.27$ eV, very close to the optimal adsorption range ($|\Delta G_{H^{*}}| < 0.20$ eV) respectively [REF]. For the other cases, Se sites shown weak binding on both structures. On $Pt_{2}HgSe_{3}$, $Pt_{Hg}$ showed the second best $\Delta G_{H^{*}}$ followed by Se and $Pt_{Se}$. On the other hand, $Pt_{Zn}$ site showed the weakest binding strength. On both $Pt_{Hg/Zn}$ sites, the H intermediate migrates to the neighboring site, as confirmed by the differential charge analysis shown in the supporting information. The differential charge analysis for the Hg and Zn sites are presented in Figure \ref{FIG:2}.f. Zn sites show more charge accumulation and less charge rearrangement in the neighboring Pt and Se atoms compared to the Hg.

\begin{table}[h!]
\caption{H adsorption free energies}\label{tab:delta-G}
	\centering
	\begin{tabular}{cc}
        Site & $\Delta G_{H^{*}}$ (\AA) \\ \hline
        $Pt_{2}HgSe_{3}(Pt)$ & 0.89 \\
        $Pt_{2}HgSe_{3}(Pt_{Hg})$ & 1.00 \\
        $Pt_{2}HgSe_{3}(Se)$ & 0.66 \\
        $Pt_{2}HgSe_{3}(Hg)$ & 0.25 \\ \hline
        $Pt_{2}ZnSe_{3}(Pt)$ & 2.47 \\
        $Pt_{2}ZnSe_{3}(Pt_{Zn})$ & 2.21 \\
        $Pt_{2}ZnSe_{3}(Se)$ & 3.43 \\
        $Pt_{2}ZnSe_{3}(Zn)$ & 0.27 \\
        \end{tabular}
\end{table}


The results presented thus far suggest that replacing Hg with Zn retains most of the material’s electronic and catalytic characteristics. However, structural properties—such as lattice parameters and buckling height—differ, which is expected due to the distinct atomic properties of each element, including atomic and van der Waals radii, as well as electronic configuration (Hg contains 4f electrons, while Zn has only 3d). Nevertheless, the similarity in key properties supports the substitution with Zn, which poses less environmental harm compared to Hg.

\begin{figure}[t!]
\centering
\includegraphics[scale=0.7]{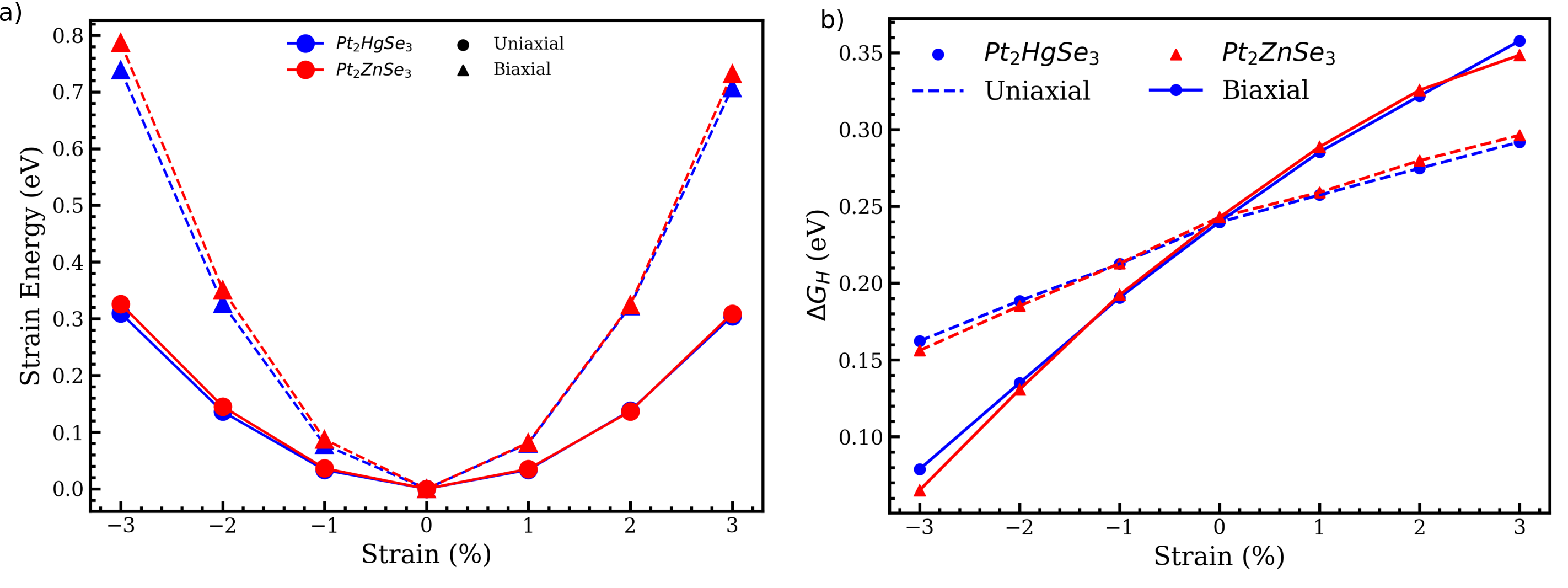}
 \caption{a) Strain energy for uniaxial and biaxial strain. The quadratic behavior of the data corroborates the elastic regime to 3\% of (compressive) tensile strain. b) The H adsorption free-energy ($\Delta G_{H^*}$) as a function of strain.}
\label{FIG:3}
\end{figure}

In the following, we investigate the effects of uniaxial and biaxial strain on the catalytic properties of the $Pt_{2}XSe_{3}$ structures. The strain ($\varepsilon$) is defined as:
\begin{align}
    \varepsilon = \frac{(a - a_{0})}{a_{0}}\times 100 \%
\end{align}
where $a_{0}$ represents the equilibrium lattice constant and $a$ the deformed lattice constant. For each value of strain, the atoms in the structure are allowed to relax with fixed lattice constraints. Uniaxial strain was applied considering deformations along the $a$ lattice vector only, while biaxial strain was achieved by simultaneously stretching/compressing $a$ and $b$ lattice vectors. To ensure that the deformations on the structures are in the elastic regime, we calculate the strain energy ($E_{s}$) defined as\cite{Yang2025}:
\begin{align}
    E_{s} = E(\varepsilon) - E_{0}
\end{align}
where $E(\varepsilon)$ represent the total energy of the strained system and $E_{0}$ the total energy of the unstrained structures. As shown in Figure \ref{FIG:3}.a, (uni/bi)axial deformations within $3\%$ (in steps of $1 \%$) result in a quadratic behavior for $E_{s}$, indicating that deformations within this range are reversible and therefore the elastic regime is conserved \cite{Yang2025}. The results for the $\Delta G_{H^{*}}$ as a function of $\varepsilon$ are presented in Figure \ref{FIG:3}.b. The results shows that the adsorption strength of the intermediate can be effectively modulated when the structure is under elastic deformation. For instance, tensile strain weakens the interaction, resulting in an upshift in the $\Delta G_{H^{*}}$. On the other hand, compressive strain results in stronger adsorption, modulating the $\Delta G_{H^{*}}$ towards zero. Moreover, it can be seen that biaxial strain is much more effective in modulating the adsorption strength, for instance at 3\% uniaxial compressive strain the $\Delta G_{H^{*}}$ for the Hg and Zn sites are 0.16 and 0.15 whereas the same percentage of biaxial strain results in 0.07 and 0.06 eV. The presented results clearly show that biaxial strain can effectively modulate the binding strength of H intermediates, thus improving the catalytic activity of the structures. Interestingly, the observed trends are in contrast to those reported in early transition metals: in $MoS_{2}$, compressive strain is found to lower the binding strength \cite{Li2016}. On the other hand, the opposite trend was also reported for Cu, which showed decreased (increased) activity for tensile(compressive) strain, consistent with our observations \cite{YanStrainOnNiPtCu2016}. The mechanisms underlying the activity changes are related to the electronic structure modifications that occur with elastic deformations and will be discussed in the following.


\begin{figure}[t!]
\centering
\includegraphics[scale=0.625]{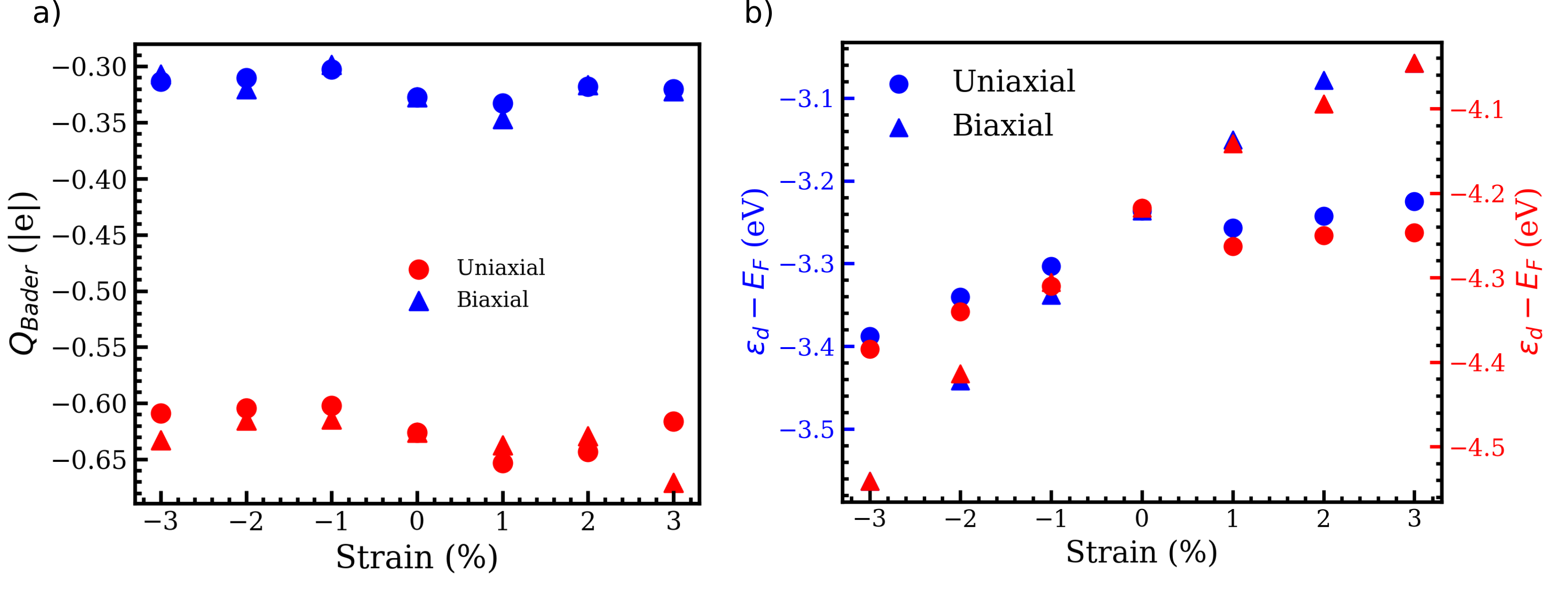} 
 \caption{a) Bader charge on the adsorbed H, b) Position of d-band center relative to the Fermi level ($\varepsilon_{d} - E_{F}$).}
\label{FIG:4}
\end{figure}

The Bader charges on the metal site (X = Hg or Zn) as a function of strain were calculated and are shown in Figure \ref{FIG:4}.a. The overall behavior is the same when comparing uniaxial and biaxial strain. The charge on the metal site decreases (minimally) as the structure is compressed, depicting more charge transfer. This is observed for both types of strain, although in $Pt_{2}ZnSe_{3}$, uniaxial strain causes a charge decrease for -1\% of compressive strain, but returning to the initial charge state as the structure is further compressed to -3\%. However, tensile deformation does not show a clear trend. For $Pt_{2}HgSe_{3}$ the charge seems to fluctuate around the unstrained structure values for both types of strain. In $Pt_{2}HgSe_{3}$ a similar behavior is observed until 2\% tensile strain; however, at 3\% uni and biaxial strain results in opposite trends, the first decreases whereas the latter increases the charge. The modulation of $\Delta G_{H^{*}}$ as a function of strain has been previously reported and attributed to the optimization of the d-band center ($\varepsilon_{d}$) in transition metal-based compounds \cite{JeonPtS2strain2025}. This theoretical descriptor was proposed to correlate the adsorption strength of adsorbates with the electronic states arising from the hybridization of the catalyst d orbitals and s or p states from adsorbate molecules \cite{HAMMER1995211}. In its simplest form, the $\epsilon_{d}$ is calculated as:
\begin{align}
    \epsilon_{d} = \frac{\int_{\infty}^{E_{F}}ED(E)dE}{\int_{\infty}^{E_{F}}D(E)dE}
\end{align}
where $D(E)$ represents the d-band density of states at the $E$ energy eigenvalue. 

\begin{figure}[t!]
\centering
\includegraphics[scale=0.625]{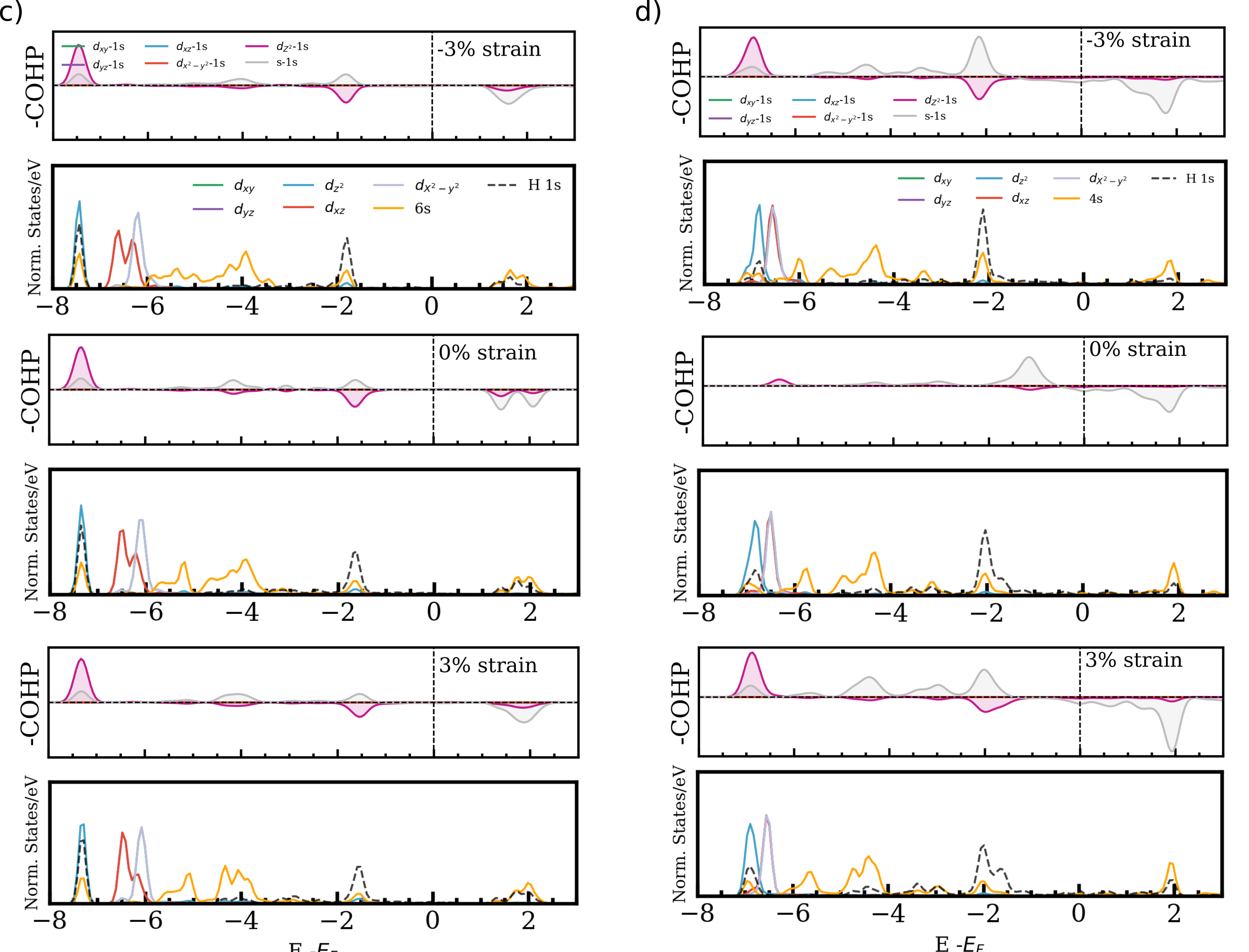} 
 \caption{The Partial Crystal Orbital Hamilton Population (pCOHP) 
 aligned with projected Density of States as a function of strain for a) Hg and b) Zn..}
\label{FIG:5}
\end{figure}

The position of the d-band center relative to the Fermi level ($\varepsilon_{d} - E_{F}$) for the metal sites as a function of the strain is presented in Figure \ref{FIG:4}. Interestingly, the observed trends are well correlated to the $\Delta G_{H^{*}}$: compressive strain lowers the $\varepsilon_{d}$ (becomes further from $E_{F}$), while tensile strain upshifts $\varepsilon_{d}$ (becomes closer to $E_{F}$). An interesting result is that the calculated values for $\varepsilon_{d} - E_{F}$ for Hg (left y axis in Figure \ref{FIG:4}.b) and Zn (right y axis on Figure \ref{FIG:4}.b) differ by exactly 1 eV. The modulation of the $\varepsilon$ with strain is well-known in literature \cite{SchnurGrossPBR2010}: for late transition metals (more than half-filled d-bands), the width of the d-band becomes narrower (wider) as the structure is compressed (stretched) and to maintain the same level of filling, $\epsilon_{d}$ shifts downwards (upwards) resulting in the observed trend. To further explore the changes in the electronic structure, we analyzed the projected Density of States on the 5d (3d) and 6s (4s) orbitals of Hg (Zn) and H 1s states and the projected Crystal Orbital Hamilton population for the Hg/Zn and H pairs (Figures \ref{FIG:5}.a, b). For Hg at 0\% strain, there is a superposition between 5$d_{z^{2}}$, 6s orbitals with H 1s at -8 and -1.75 eV below $E_{F}$. On the other hand, for Zn these peaks are located at -7 and -2 eV below $E_{F}$. The significant overlap between these orbitals indicates the hybridization between them. The overlap at lower energies contributes to the formation of anti-bonding states, whereas the hybridization closer to the Fermi level produce a bonding orbital for both Hg and Zn..

Lattice expansion causes depletion of the prominent peak at -2 eV below $E_{F}$, close to the bonding peak formed by hybridization between Hg 6$s$ and H 1$s$, as shown by the pCOHP (Figure SX). The same trend is observed for Zn (Figure 4.d), further corroborating the equivalence between the electronic and catalytic properties of these two elements. The depletion in the 6$s$ (4$s$) orbitals indicates the decrease in the electron population in weakening the binding strength.


\section{Conclusions}

In this work, we have conducted state-of-art ab-initio simulations based on DFT to investigate the catalytic properties of $Pt_{2}XSe{3}$ (X = Hg, Zn) in HERs. The substitution of Hg by Zn results in the same overall electronic and catalytic behavior, given the similar electronic distribution of both elements. The late transition metal sites exhibit the best activity of HER at pH = 0, U = 0 at the SHE scale. Strain can modulate the H adsorption free energy ($\Delta G_{H^{*}}$) towards zero, achieving almost thermoneutral H binding ($\Delta G_{H^{*}}$ = 0.08 and 0.07 eV for Hg and Zn, respectively) at 3\% compressive strain. Bader charge analysis revealed no obvious trend in the charge of the metal sites as a function of the strain, indicating that the origin of the modulation in the intermediate binding strength is related to other factors. Tensile (compressive) strain downshifts (upshifts) the d-band center of the late transition metal sites, unveiling an interesting pattern that resembles the $\Delta G_{H^{*}}$, characteristic in transition metals with more than half-filled d orbitals. Furthermore, the pDOS and pCOHP analysis revealed that lattice stretching results in the depletion of charge population on hybrid states corresponding to bonding states, contributing to the destabilization of the H-metal bonds. Our contribution explores strain engineering as an effective strategy to tailor the activity of 2D mineral-based catalysts for HER, advancing our understanding of how mechanical manipulation can effectively modulate the catalytic properties of these materials. 

\begin{suppinfo}

\begin{itemize}
    \item Data available at the supplementary material;
\end{itemize}

\end{suppinfo}

\begin{acknowledgement}

This work was funded by São Paulo Research Foundation (FAPESP - process number \#2024/11376-6), Coordenação de Aperfeiçoamento de Pessoal de Nível Superior (CAPES finance code 001) and CNPq (Process 308428/2022-6). The authors also acknowledge the UFABC Computation Multiuser Center (CCM) for the computational resources provided.

\end{acknowledgement}




\bibliography{achemso-demo}

\end{document}